# Ultrahigh photoresponse of few-layer $TiS_3$ nanoribbon transistors

*Joshua O. Island*[1,]*, *Michele Buscema*[1], *Mariam Barawi*[2], *José M. Clamagirand*[2], *José R. Ares*[2], *Carlos Sánchez*[2], *Isabel J. Ferrer*[2], *Gary A. Steele*[1], *Herre S.J. van der Zant*[1], *and Andres Castellanos-Gomez*[1,]*

[1] Kavli Institute of Nanoscience, Delft University of Technology, Lorentzweg 1, 2628 CJ Delft, The Netherlands.
[2] Materials of Interest in Renewable Energies Group (MIRE Group), Dpto. de Física de Materiales, Universidad Autónoma de Madrid, UAM, 28049- Madrid, Spain.

E-mail: j.o.island@tudelft.nl , a.castellanosgomez@tudelft.nl

Transition metal chalcogenides have raised a huge interest in the nanoscience and material science communities.[1-3] The possibility of isolating ultrathin layers of these materials opens the door to new applications and phenomena derived from the reduced dimensionality. Among the large family of semiconducting chalcogenides, Mo- and W- based dichalcogenides are the most studied materials because of their electronic and optical properties which could make them complementary materials to graphene in applications requiring optically active semiconducting materials.[4-14] Nonetheless, there are many other semiconducting chalcogenide materials where electronic properties, in atomically thin form, are thus far unexplored.

$TiS_3$, for instance, is one of the semiconducting members of the trichalcogenides family with a bulk optical bandgap of ≈ 1 eV.[15] It has a structure, shown in **Figure 1(a)**, composed of parallel sheets of one dimensional chains of stacked triangular prisms ($TiS_3$).[16] The sheets are held together by van der Waals forces and might be exfoliated in the same manner as graphene and other layered chalcogenide materials.[3, 17] Although the bulk electronic properties have been studied in fair detail,[18-21] little is known of the properties of nanostructured $TiS_3$ which is a promising material for optoelectronic applications. While Mo- and W- based dichalcogenides present a direct bandgap only at the single-layer limit, macroscopic films of $TiS_3$ ribbons (with thicknesses of hundreds of nanometers) show a direct bandgap of 1.1 eV and photocurrent response to white light illumination.[22, 23]

Here, we isolate individual exfoliated, few-layer $TiS_3$ nanoribbons to fabricate field effect transistors (NR-FET). The electrical characteristics are studied finding an n-type conduction behavior with current on/off ratios as high as $10^4$ and mobilities up to 2.6 $cm^2$ $V^{-1}$ $s^{-1}$. The photoresponse of the fabricated NR-FETs is studied at different illumination wavelengths and powers. The fabricated devices show an ultrahigh photoresponse up to 2910 A $W^{-1}$ and they respond to wavelengths across the visible spectrum. In addition, these devices present rise/fall times of about 4/9 ms and a $f_{3dB}$ frequency up to 1000 Hz. The excellent combination of FET characteristics, large photoresponse and high cutoff frequency make atomically thin $TiS_3$ NR-FETs superior to state-of-the-art phototransistors based on other semiconducting 2D materials.[24]

$TiS_3$ NRs are grown by sulfuration of bulk Ti disks at 500 °C.[23] The morphology and stoichiometry of the sample can be accurately controlled by adjusting the process temperature and sulfur pressure. The growth results in gram-scale quantities of NR film which may be useful for applications such as a filler for nanocomposites. The resulting NR film is characterized by electron microscopy and X-ray diffraction. Figure 1(b) shows a bright field transmission electron microscopy (TEM) image of a single grown $TiS_3$ ribbon. The crystal planes are determined from the electron diffraction pattern (inset). The growth direction of the NRs is along the b-axis. Scanning electron microscopy (SEM) images of the film and a single NR are shown in Figure 1(c) and Figure 1(d), respectively. As grown NRs have lengths of hundreds of microns, widths of a few microns, and thicknesses of hundreds of nanometers. Further characterisation in the form of X-ray diffraction (XRD) and Raman spectroscopy can be found in the Supporting Information (Figures S1 and S2, respectively). To fabricate single ribbon devices, $TiS_3$ ribbons are transferred from the bulk disks onto degenerately doped silicon substrates with 285 nm of thermal $SiO_2$ using an all-dry viscoelastic stamping method.[25] We find that during the transfer, the $TiS_3$ nanoribbons are mechanically exfoliated leading to a decreased thickness with respect to the as-grown ribbons. Ribbons with a thickness of 10-30 nm can be easily selected due to their distinct contrast under an optical microscope to fabricate FETs (see Supporting Information). Standard electron beam lithography and metal evaporation are used to deposit Au electrodes (50 nm) with a titanium sticking layer (5 nm). Finally, fabricated samples are annealed at 250 °C for 2 hours in a mixture of argon and hydrogen (5:1). Characteristics of 10 fabricated and measured devices can be found in the Supporting Information (Table S2) but for consistency all figures of the main text have been obtained from a single device.

In **Figure 2** we show the room temperature electrical characterisation of a fabricated NR-FET device. An AFM micrograph of the device is presented in Figure 2(a). A line profile taken across the NR shows a thickness of 22 nm and a width of 195 nm. Measurements are made between two adjacent electrodes of the full device. Figure 2(b) plots the source-drain current as a function of the bias voltage ($I$-$V_b$) at different back-gate voltages ($V_g$). The $I$-$V_b$ characteristics show a strong dependence on the back-voltage, are symmetric with bias voltage, and show linear behavior. This suggests that we can switch the device ON with the gate voltage and that the Ti/Au electrodes provide good contact to the $TiS_3$ NR. Figure 2(c) shows the source-drain current versus back gate voltage ($I$-$V_g$) at constant bias voltages. The device has a clear n-type behavior. At $V_b$ = 1V, for negative $V_g$ the current through the device is ~ 1.5 nA in the OFF state while for positive $V_g$, the current reaches ~1.1 µA in the ON state giving an ON-OFF ratio of ~700. Using the $I$-$V_g$ data of this device we calculate the transconductance ($\partial I/\partial V_g$) to estimate the mobility from

$$\mu = \frac{L}{ZC_iV_b}\frac{\partial I}{\partial V_g} \tag{1}$$

where $L$ is the channel length, $Z$ is the width of the nanoribbon, and $C_i$ is the oxide capacitance per unit area.[26] The capacitance is calculated using finite element methods with COMSOL MULTIPHYSICS v4.3b to account for fringing effects derived from the reduced width of the $TiS_3$ ribbon channel (comparable to the dielectric thickness).[27] From Equation 1 we estimate a carrier mobility of 1.4 $cm^2\ V^{-1}\ s^{-1}$ (up to 2.6 $cm^2\ V^{-1}\ s^{-1}$ for other devices, see Supporting Information) which is an order of magnitude lower than the measured bulk value of 30 $cm^2\ V^{-1}\ s^{-1}$.[18] Note that this estimate is a lower bound as it is obtained from a two terminal measurement without correcting for the contact resistance (a correction which typically yields higher mobility values). While the mobility estimates are modest, they compare well with recently fabricated $MoS_2$ photodetectors with mobilities of 4 $cm^2\ V^{-1}\ s^{-1}$ and lower and may be improved by optimizing device geometry and dielectric material.[5, 6]

The photoresponse of the $TiS_3$ NR devices is studied by measuring their electrical characteristics upon illumination. The optoelectronic measurements are carried out in a vacuum probe station equipped with 5 lasers (λ = 532 nm, 640 nm, 808 nm, 885 nm, and 940 nm). **Figure 3** shows the optoelectronic characterisation of the same NR-FET device shown in Figure 2. Figure 3(a) shows $I$-$V_g$ characteristics recorded at $V_b$ = 500 mV in dark (solid black line) and under 532 nm laser excitation (P = 500 μW, solid green line). There is a steady increase in current over the entire gate voltage range, reaching a photocurrent ($I_{ph}$, difference between current under illumination and in dark) of 100 nA at $V_g$ = +40 V. The inset shows the $I$-$V_b$ characteristics ($V_g$ = -40 V) recorded at the same power and wavelength. In the OFF state the photocurrent reaches 40 nA. In Figure 3(b) we plot the $I$-$V_b$ characteristics ($V_g$ = -40 V) for different excitation powers of a 640 nm laser showing a clear increase in the conductance of the NR-FET with increasing power. The corresponding photocurrent, plotted in Figure 3(c), increases sublinearly with respect to power, $I_{ph} \propto P^{\alpha}$ where α = 0.7, and points to a complex process of photocarrier generation, recombination, and trapping.[28, 29] We attribute photocurrent generation in these devices primarily to photovoltaic effects and not to the photo- thermoelectric effect recently shown for single-layer $MoS_2$ photodetectors.[30] In order to generate photocurrent through the photo-thermoelectric effect a sizeable thermal gradient is required across the device (usually achieved by employing a very focalized laser spot heating one electrode).[30] In our case, we employ a large size laser spot (much larger than the device area) and we thus expect a homogeneous temperature all across the device. Moreover, the photo-thermoelectric effect presents itself as photogenerated current even in the absence of bias voltage.[30] In our measurements, however, there is no photocurrent generated at zero bias within experimental resolution (inset of Figure 3(a)). In addition, we find that the responsivity decreases as the density of charge carriers is decreased with the gate voltage (see Supporting Information) contrary to what is expected for photo-thermoelectric generated current.[30]

To characterize the optical gain of the device, we calculate the responsivity from the photocurrent, $Res = \frac{I_{ph}}{P}$, where $P$ is the power of the laser which has been scaled by the ratio of the area of the device (195 x 470 nm) to the area of the laser spot (diameter 200 μm). Figure 3(d) plots the responsivity as a function of the laser power. The responsivity reaches 1030 A $W^{-1}$ at a power of 100 nW and decreases with increasing power (2910 A $W^{-1}$ for the best device, see Supporting Information). This responsivity is several orders of magnitude larger than intrinsic graphene photodetectors and surprisingly larger than the most recent $MoS_2$ photodetectors which achieve a responsivity up to 880 A $W^{-1}$.[5, 31] ZnO nanowire devices show responsivities of $10^4$ A $W^{-1}$ but unlike $TiS_3$ NR-FETs, they operate outside the visible range of the spectrum.[32] Responsivities of $TiS_3$ devices may be further improved by fabricating hybrid $TiS_3$-quantum dot devices as already demonstrated for graphene based photodetectors.[33] The decrease in responsivity with increased power could be attributed to filling of the long-lived trap states at higher powers which are responsible for high photoconductive gain.[34, 35] Using the responsivity of 1030 A/W, we estimate the external quantum efficiency (EQE) by $EQE = R(hc)/(e\lambda)$. The EQE for this device at wavelength of 640 nm reaches 2 x $10^5$ % and is larger than the recent high gain $In_2Se_3$ nanosheet devices with an EQE of 1.6 x $10^5$ %.[36] In Figure 3(e) we plot the photocurrent at $V_b$ = 1V and $V_g$ = -40 V for different excitation wavelengths at a power of 500 μW. By linearly extrapolating the photocurrent vs. excitation wavelength data, we extract a rough estimate of the bandgap energy of 1.2 eV (1010 nm) in close agreement with previous measurements on macroscopic $TiS_3$ thin films having a direct bandgap of 1.1 eV.[22] The EQE calculated for each wavelength can be found in the Supporting Information.

The frequency response of the $TiS_3$ NR-FETs is characterized by modulating the intensity of the excitation laser. In **Figure 4**, we show the photo-switching characteristics of the same NR-FET by modulating the 640 nm (P = 500 μW) laser excitation with a mechanical chopper at $V_g$ = -40 V and $V_b$ = 1 V. Reliable, repeated cycling of the NR-FET at a chopper frequency of 10 Hz is shown in Figure 4(a). In Figure 4(b), a zoom on one switching cycle is shown. The 10-90% rise time extracted is ≈4 ms and the 10-90% fall time is ≈9 ms. The slower fall time is most likely related to the relaxation of deep trap states. These time scales are an order of magnitude faster than the lowest rise/fall times measured in $MoS_2$ devices and faster than the high gain ZnO nanowire devices.[6, 32] In Figure 4(c) we plot the device responsivity for increasing modulation frequency of the 640 nm laser at a power of 500 μW. A cutoff frequency ($f_{3dB}$) of 1000 Hz is reached which is at the limit of our measurement set-up. This $f_{3dB}$ puts $TiS_3$ NR-FETs among the best nanostructured devices for photodetection.[37]

The direct bandgap (1.1 eV), high sensitivity, and fast response make $TiS_3$ an ideal material for optoelectronic applications. In Table 1 we summarize the latest high responsivity photodetectors. The $TiS_3$ NR-FETs have the best combination of responsivity and speed for visible and near-infrared light detection. The bandgap energy of 1.1 eV (cutoff wavelength of 1130 nm) makes them a suitable replacement to microstructured Si in applications where high gain is needed like night vision cameras.

In conclusion, one of the first characterisations of $TiS_3$ in nanostructured form has been performed. Few-layer $TiS_3$ NR-FETs have been fabricated and measured in dark and under illumination at room temperature. We find an n-type semiconducting behavior with mobilities of 2.6 $cm^2\ V^{-1}\ s^{-1}$ and ON/OFF ratios up to $10^4$. The NR-FETs display an ultrahigh photoresponse up to 2910 A $W^{-1}$ and fast switching times of about 4 ms with a cutoff frequency of 1000 Hz . The excellent combination of FET characteristics, high photoresponse and fast switching rates make atomically thin $TiS_3$ an interesting material for photodetection and photovoltaic applications. This work opens a new avenue to exploit the family of Ti-based trichalcogenides in 2D electronics and optoelectronics.

Experimental Section

*Characterisation of $TiS_3$ nanoribbons.* X-ray diffraction, transmission electron microscopy and scanning electron microscopy is employed to characterize the structure and topography of the as-grown $TiS_3$ NRs. We also perform Raman microscopy on the isolated exfoliated NRs to confirm their structure. Raman measurements are performed in a Renishaw *in via* system in backscattering configuration ($\lambda$ = 514 nm, 100× objective with NA = 0.95). This configuration has a typical spectral resolution of ~ 1 $cm^{-1}$. To avoid laser-induced modification or ablation of the samples, all spectra were recorded at low power levels of 100 – 500 μW. Atomic Force Microscopy (AFM) is used to measure the thickness of the NRs. The AFM (*Digital Instruments D3100 AFM*) is operated in amplitude modulation mode with Silicon cantilevers (spring constant 40 N $m^{-1}$ and tip curvature <10 nm).

*Fabrication of $TiS_3$ nanoribbon-based field effect transistors(NR-FETs):* Single NRs are isolated from an as-grown film and subsequently transferred to a pre-patterned $SiO_2$/Si substrate with an all-dry method. Thin NRs are then selected by their optical contrast. We use standard e-beam lithography and lift-off procedures to define the metallic contacts (5 nm Ti/50 nm Au).

*Characterisation of the NR-FETs*: Optoelectronic characterisation is performed in a *Lakeshore Cryogenics* probestation at room temperature in vacuum (<$10^{-5}$ mbar). The excitation is provided by diode pumped solid state lasers operated in continuous wave mode (CNI Lasers). Intensity modulation is provided by a mechanical chopper. The light is coupled into a multimode optical fiber through a parabolic mirror. At the end of the optical fiber, another identical parabolic mirror collimates the light exiting the fiber. The beam is then directed into the probe station's zoom lens system and then inside the sample space. The beam spot size on the sample has a 200 μm diameter for all wavelengths.

Acknowledgements
This work was supported by the European Union (FP7) through the program RODIN and the Dutch organization for Fundamental Research on Matter (FOM). A.C-G. acknowledges financial support through the FP7-Marie Curie Project PIEF-GA-2011-300802 ('STRENGTHNANO'). Authors from MIRE Group acknowledge the support of the Ministry of Economy and Competitiveness (MINECO) for this research (contract MAT2011-22780). They also thank Dr. Garcés from CSIC-CENIM for the support at TEM and ED, Dr. Bodega for suggestions and discussions and technical support from Mr. F. Moreno.

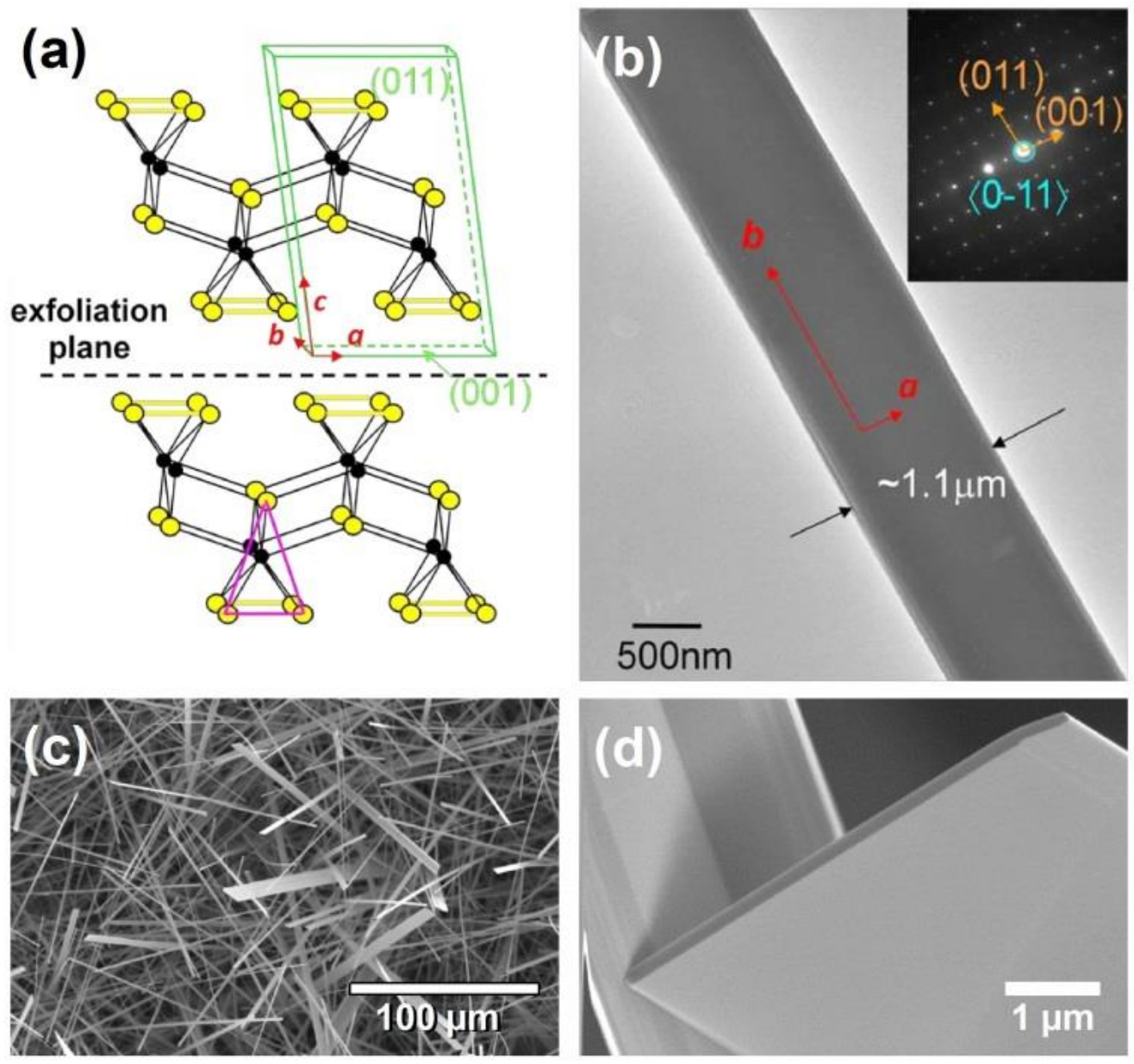

**Figure 1.** (a) $TiS_3$ nanoribbon structure composed of parallel sheets of one dimensional chains of stacked triangular prisms ($TiS_3$, base shown in pink) that are held together by van der Waals forces across the exfoliation plane. (b) Bright field TEM micrograph and the corresponding electron diffraction pattern (inset) of a $TiS_3$ nanoribbon. (c) SEM image of $TiS_3$ nanoribbon film on the Ti growth disk. (d) FEG-SEM image of the edge of one $TiS_3$ nanoribbon.

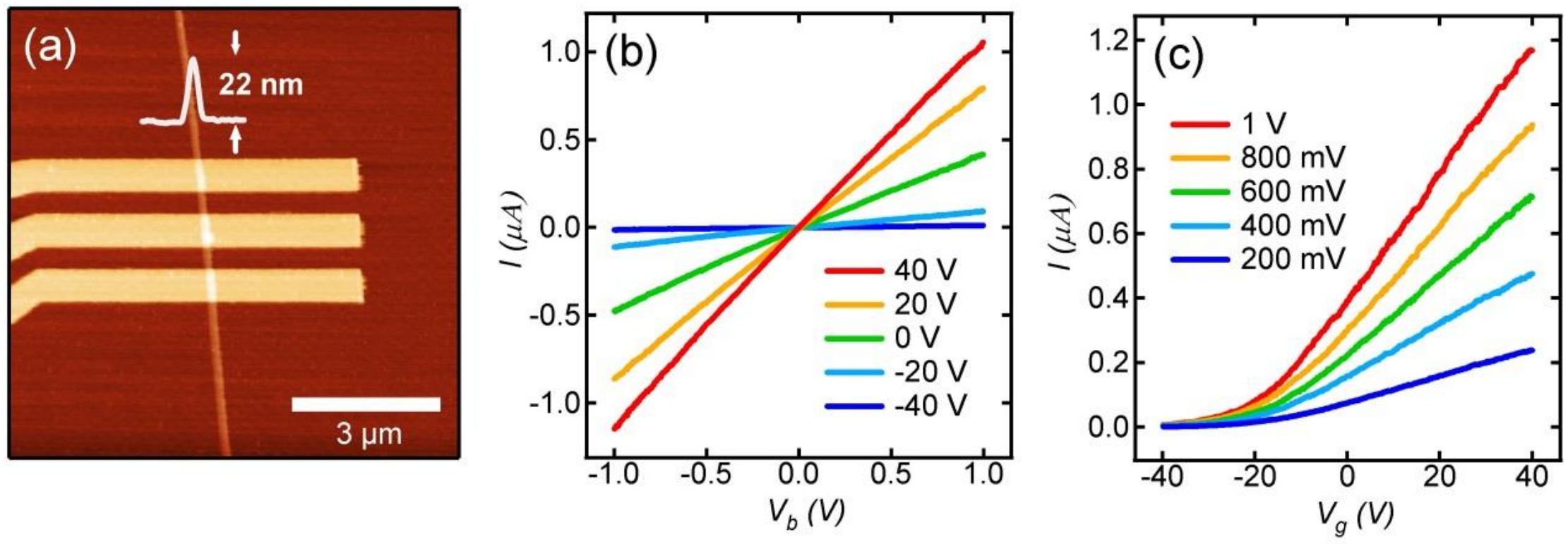

**Figure 2.** (a) AFM image of the FET device with electrodes spaced by 500 nm. The width of the $TiS_3$ NR is 195 nm. (b) Output characteristics ($I$-$V_b$) of the device at backgate voltages from -40V to +40V in steps of 20 V. (c) Transfer characteristics ($I$-$V_g$) at different bias voltages (from 200mV to 1V in steps of 200 mV).

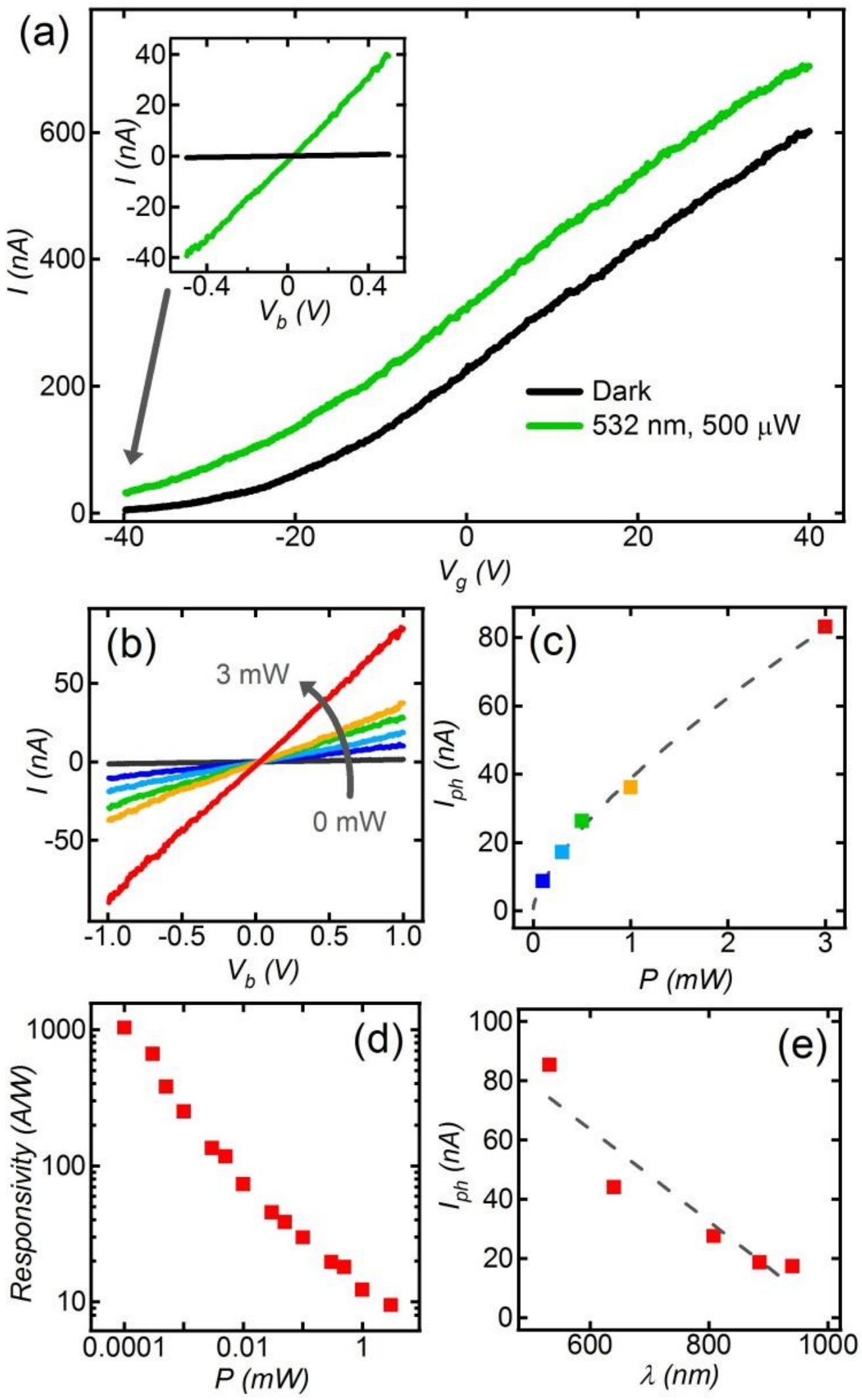

**Figure 3.** (a) $I$-$V_g$ characteristics for $V_b$ = 500 mV in dark (black solid line) and under 532 nm, 500 μW excitation (green solid line). Inset shows the $I$-$V_b$ characteristics at $V_g$ = -40 V for the same excitation. (b) $I$-$V_b$ characteristics at $V_g$ = -40 V in dark (black solid line) and under 640 nm excitation for increasing laser powers up to 3 mW. (c) Photocurrent extracted from panel (b) for increasing powers. The dashed line is a power law fit. (d) Log-log plot of the responsivity as a function of excitation power (λ = 640 nm). (e) Photocurrent measured as a function of different laser wavelengths. By linearly extrapolating (dashed line) the photocurrent vs. excitation wavelength data, we extract a rough estimate of the bandgap energy of 1.2 eV (1010 nm). Note that measurements for panels (c)-(e) are taken at $V_b$ = 1V and $V_g$ = -40 V.

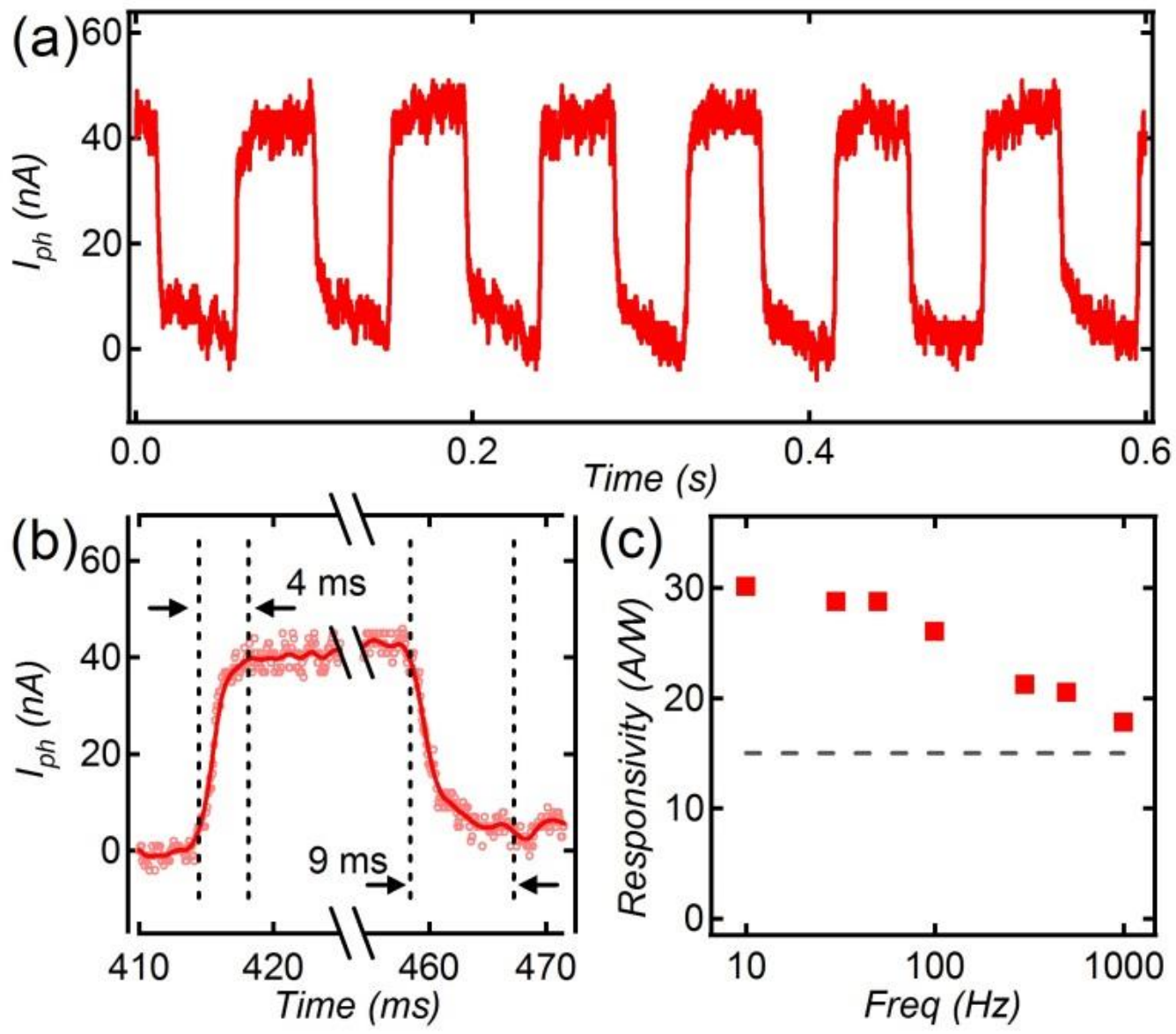

**Figure 4**. (a) Current response under a 10 Hz mechanically modulated optical excitation (λ = 640 nm, P = 500 μW). (b) Zoom on a single switching cycle at 10 Hz frequency. Rise times taken at 10% and 90% of the maximum photocurrent are ≈4 ms and fall times are ≈9 ms. The solid red curve is the smoothed result of the raw data in pink circles. (c) Responsivity versus increasing modulation frequency for an excitation wavelength of 640 nm at 500 μW. The $f_{3dB}$ for this device reaches 1000 Hz. Note that all measurements are taken at $V_b$ = 1V and $V_g$ = -40 V.

**Table 1.** Summary of recent high-responsivity photodetectors

| Material | Bandgap | Cutoff wavelength | Responsivity | Response Time | Ref. |
|---|---|---|---|---|---|
| ZnO Nanowire | ≈3.4V (direct) | 360 nm | 47000 A/W | | 32 |
| GaTe Flakes | 1.7 eV (direct) | 730 nm | 10000 A/W | 6 ms | 38 |
| $TiS_3$ NR | 1.1 eV (direct) | 1130 nm | 2910 A/W | 4 ms | This work |
| Monolayer $MoS_2$ | 1.8 eV (direct) | 690 nm | 880 A/W | 4 s | 5 |
| $In_2Se_3$ Nanosheet | 1.3 eV (direct) | 955 nm | 395 A/W | 18 ms | 36 |
| Microstructured Si | 1.1 eV (indirect) | 1130 nm | 119 A/W | | 39 |
| Multilayer GaSe | 2.11 eV (indirect) | 590 nm | 2.8 A/W | 20 ms | 40 |
| Graphene | | | 0.01 A/W | 1.5 ps | 41 |
| Multilayer InSe | 1.4 eV (indirect) | 885 nm | 34.7 mA/W | 488 μs | 42 |
| Multilayer $WS_2$ | 1.96 eV | 635 nm | 92 μA/W | 5 ms | 10 |

# Supporting Information for:

# Ultrahigh photoresponse of few-layer $TiS_3$ nanoribbon transistors

*Joshua O. Island*, Michele Buscema, Mariam Barawi, José M. Clamagirand, José R. Ares, Carlos Sánchez, Isabel J. Ferrer, Gary A. Steele, Herre S.J. van der Zant, and Andres Castellanos-Gomez**

**Supporting Information Contents**

## 1. XRD spectra of $TiS_3$ nanoribbon film after growth

X-ray diffraction (XRD) patterns were taken in a Panalytical X'pert Pro X-ray diffractometer with CuKα radiation (λ=1.5406 Å) in ϑ-2ϑ configuration. Deposits are formed by monoclinic $TiS_3$ as a unique crystalline phase, identified from the XRD diffraction pattern shown in Figure S1, which is consistent with JCPDS-ICDD 15-0783.[1] All diffraction peaks are narrow confirming that the material is well crystallized. The most intense peak corresponds to the ⟨012⟩ direction in good agreement with the reference but the relative intensity, in comparison with the other ones, is about six-fold higher, hints strong preferred orientation in this direction. From the half height wide of ⟨012⟩ peak, an average value of the crystallite size of 80(±10) nm is estimated by using the Scherrer formula.[2] Lattice parameters (*a*, *b*, *c* and *β*), determined from XRD patterns are listed in Table 1 besides those previously reported. As can be observed, our values of the unit cell dimensions are smaller than reported ones except in the *c* direction. At last, the lattice volume is smaller than reported values suggesting that the lattice of as grown $TiS_3$ is contracted (aprox. 4%) in comparison with the published one.

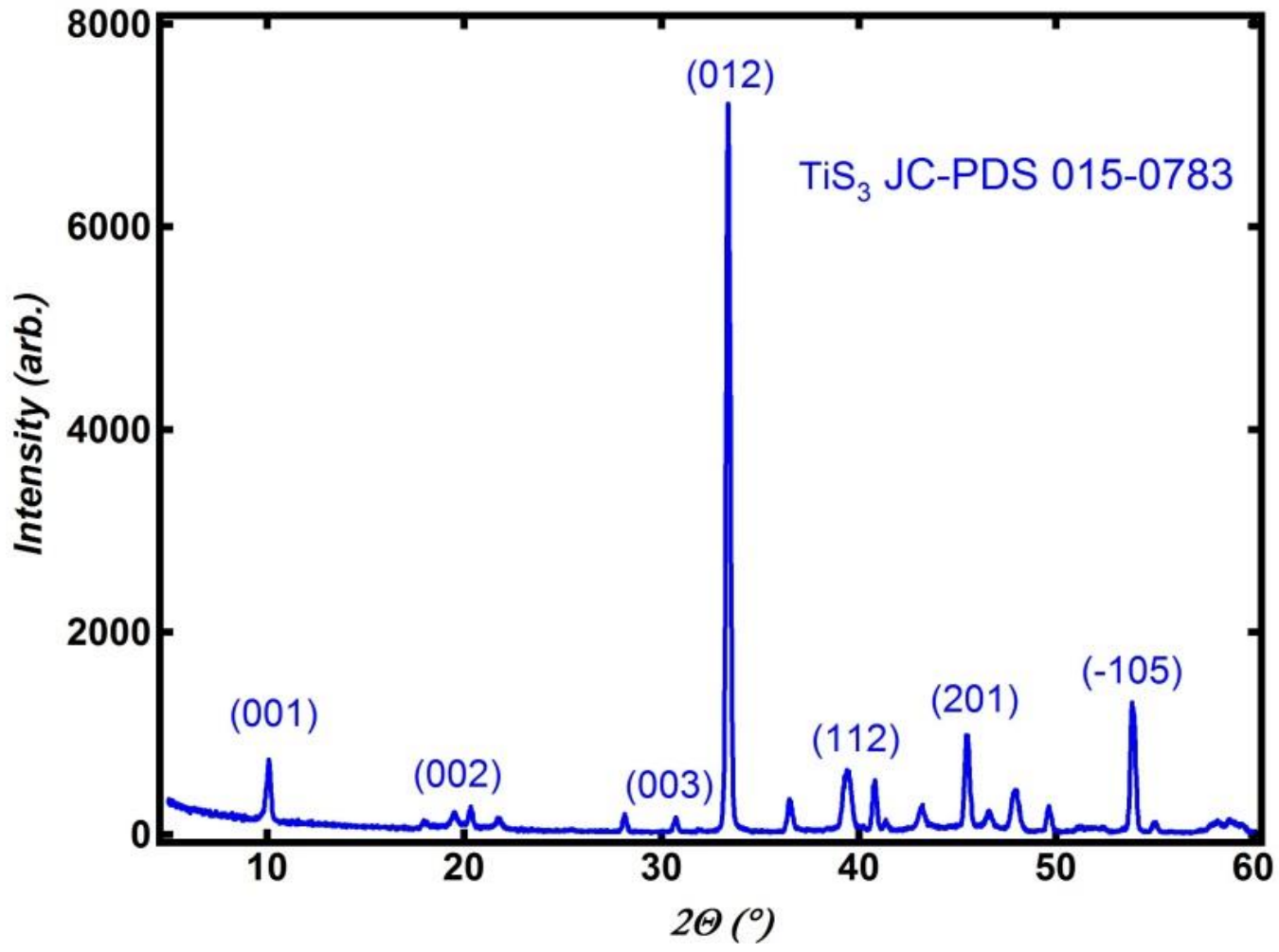

**Figure S1:** XRD patterns of $TiS_3$ formed by nanoribbons. Some Miller indices for the monoclinic structure (JCPDS-ICDD 15-0783) are included.

**Table S1:** Lattice parameters of $TiS_3$ obtained from XRD patterns and the values previously reported in Ref. 1

| | JCPDS[1] | This work |
|---|---|---|
| ***a* (Å)** | 4.973(0) | 4.947(5) |
| ***b*(Å)** | 3.433(0) | 3.397(5) |
| ***c* (Å)** | 8.714(0) | 8:780(5) |
| ***β* (°)** | 97.74 0) | 95.45(5) |
| ***V* (Å$^3$)** | 147.41(0) | 146.88 (3) |
| ***D*(nm)** | ---- | 80 ± 10 |

## 2. Raman spectra of $TiS_3$ nanoribbons after exfoliation

Raman spectra were taken on selected nanoribbons after transfer and exfoliation to confirm composition (see Figure S2). The spectra reveal three peaks corresponding to the three $A_g$-type modes of $TiS_3$.[3]

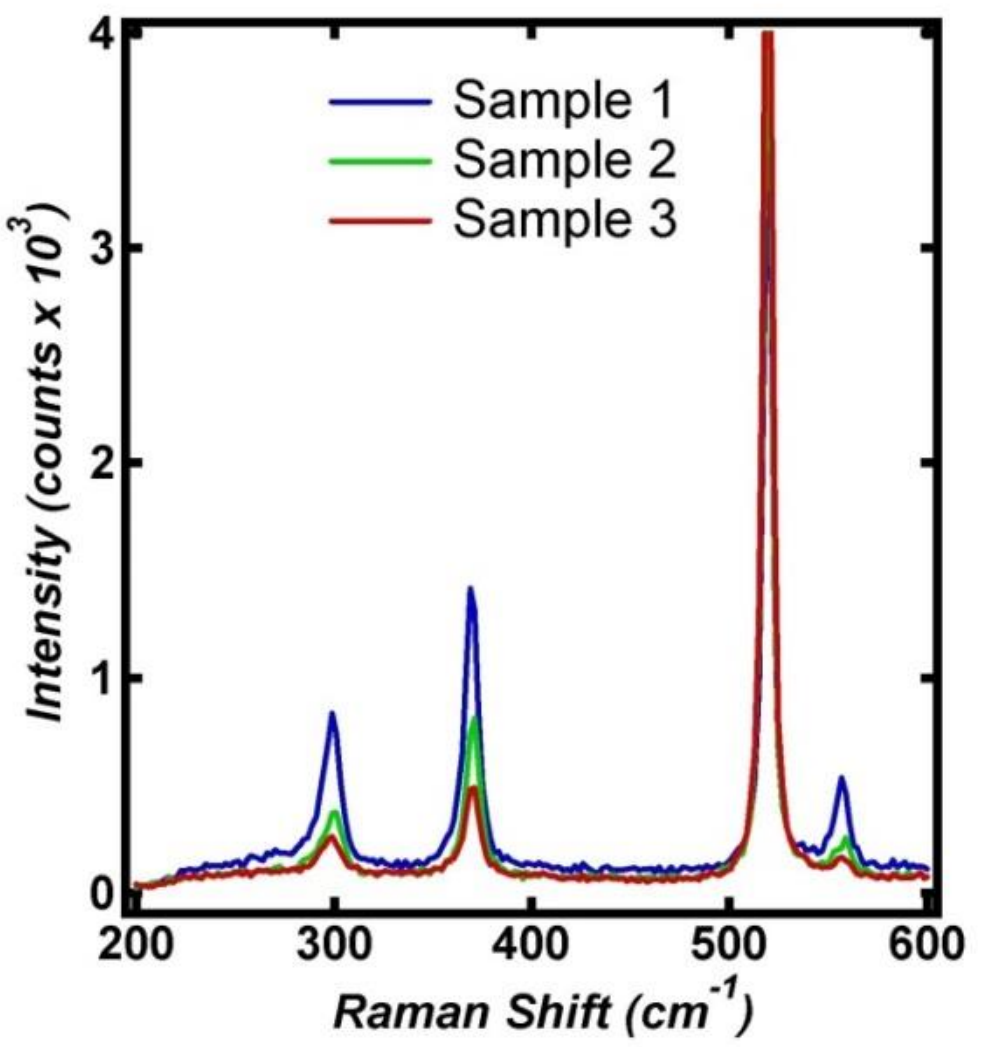

**Figure S2**: Raman spectra of three different nanoribbons showing three peaks at 300, 370, and 557 $cm^{-1}$ corresponding to $A_g$-type modes. 520 $cm^{-1}$ is the silicon substrate peak.

## 3. Optical and AFM images of exfoliated $TiS_3$ nanoribbon after transfer

An optical image of an exfoliated nanoribbon after transfer is shown in Figure S3(a). The thicker as-grown nanoribbons are clearly visible and have a light green color under an optical microscope. The exfoliated ribbon has a lower contrast but is still easily located below the thicker ribbons. An AFM image of the exfoliated ribbon is shown in Figure S3(b). The line profile taken across the nanoribbon gives a height of 13 nm.

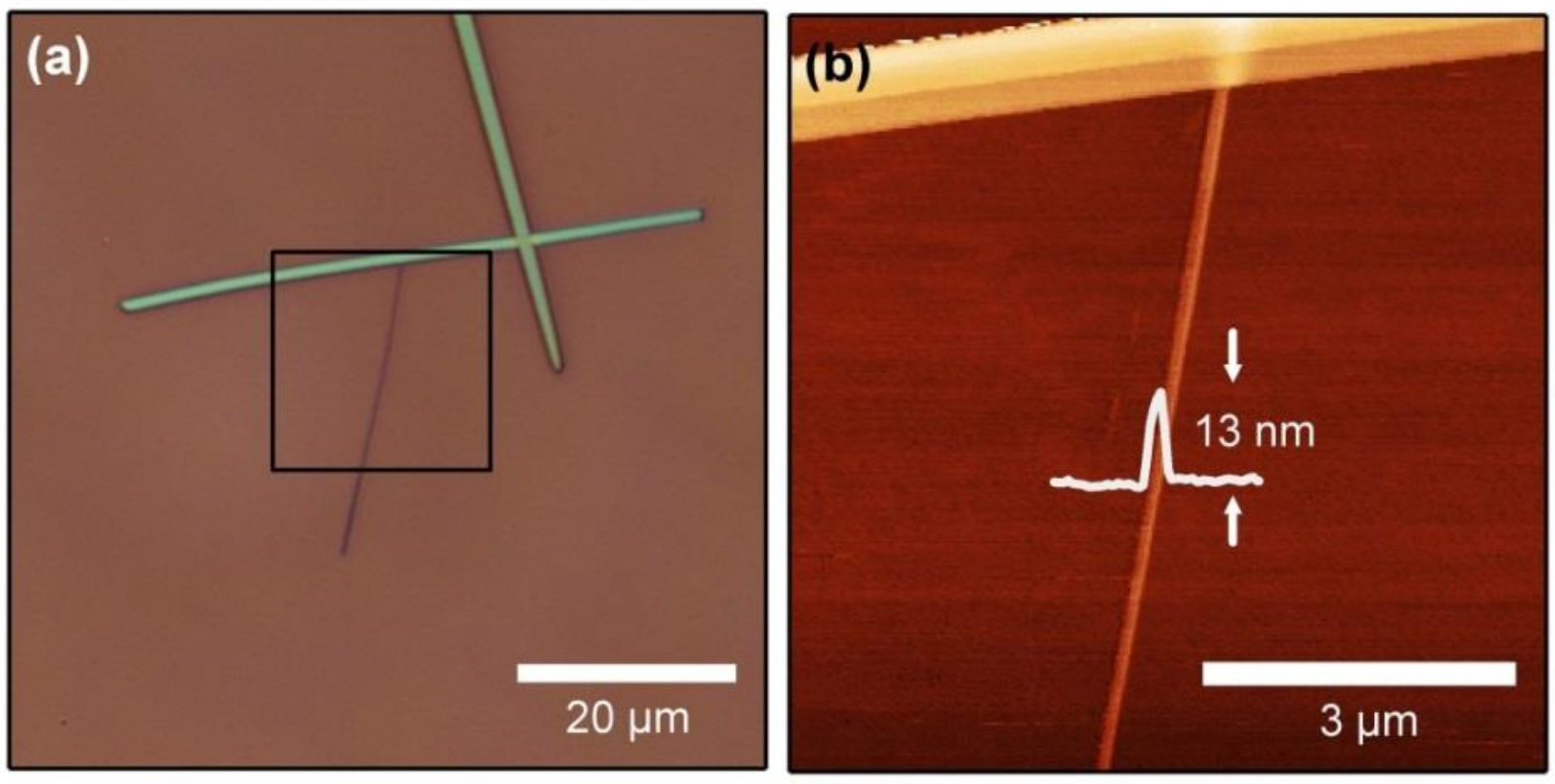

**Figure S3**: Optical image of transferred and exfoliated $TiS_3$ nanoribbon. The thicker unexfoliated nanoribbons can be seen (light green) next to the thinner exfoliated nanoribbon. The box shows the region of the AFM image in (b). (b) AFM image of the exfoliated nanoribbon with a thickness profile.

## 4. Summary of all measured $TiS_3$ nanoribbon phototransistors

In Figure S4 we show AFM images and $I$-$V_b$, $I$-$V_g$ characteristics for three additional $TiS_3$ phototransistors. Table S2 shows a summary of all measured $TiS_3$ phototransistors. The first device (JI7B1, Figure S4(a), (b), and (c)) was not annealed and the responsivity reflects the lower photoresponse (see Table S2). Subsequent devices were annealed and showed stronger photoresponse.

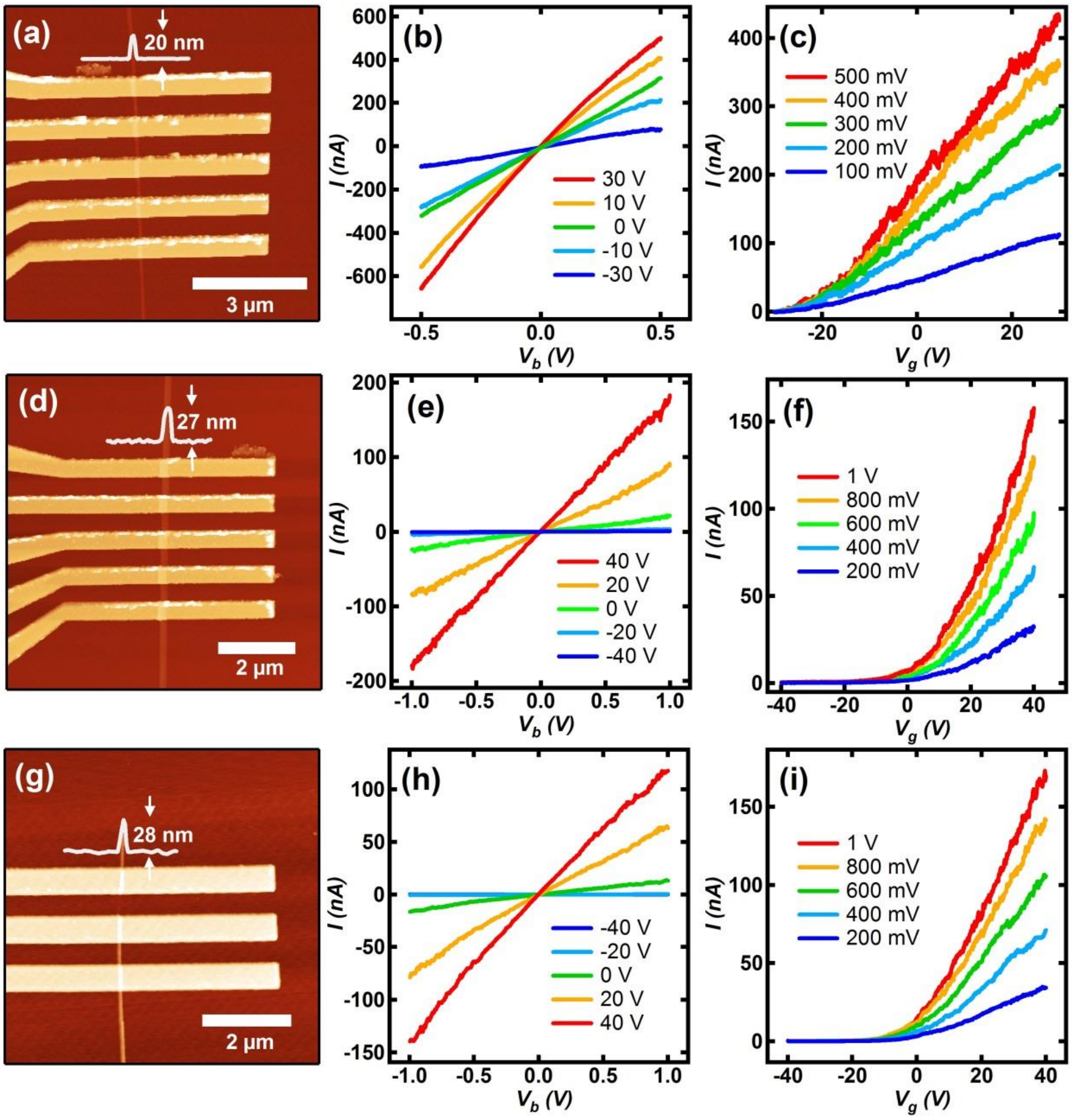

**Figure S3**: (a-c) AFM and $I$-$V_b$, $I$-$V_g$ for device JI7B1. (d-f) AFM and $I$-$V_b$, $I$-$V_g$ for device JI7B4. (g-i) AFM and $I$-$V_b$, $I$-$V_g$ for device JI9B5.

**Table S2:** Summary of all measured NR-FET devices.

| Device | Thickness [nm] | ON/OFF ratio | Mobility [$cm^2 V^{-1} s^{-1}$] | Responsivity [$A W^{-1}$] | Rise/fall time[a] [ms] | $f_{3dB}$ [Hz] |
|---|---|---|---|---|---|---|
| JI7B1[b] | 20 | 850 | 2.6 | 320 | --- | --- |
| JI7B4 | 28 | 315 | 0.5 | 660 | 12/15 | 500 |
| JI9A2[c] | 22 | 730 | 1.4 | 1030 | 4/9 | 1000 |
| JI9A3 | 15 | 400 | 1.6 | --- | --- | --- |
| JI9A4 | 36 | 90 | 1.3 | 530 | --- | --- |
| JI9A5 | 15 | 75 | 2.0 | 2590 | --- | --- |
| JI9A6 | 19 | 725 | 0.8 | 430 | --- | --- |
| JI9B3 | 22 | 740 | 1.5 | 2910 | --- | --- |
| JI9B5 | 27 | 12300 | 0.5 | 110 | --- | --- |

[a)] Times extracted between 10% and 90% of the maximum photocurrent
[b)]This device was not annealed; [c)] Main text device

## 5. External quantum efficiency (EQE) for different wavelengths

Table S3 presents a complete responsivity dataset acquired for different wavelengths with laser excitation power of 500 µW ($V_b$ = 1V and $V_g$ = -40 V). The EQE values are estimated for the different wavelengths from these numbers. Note that a higher responsivity would be obtained for even lower laser excitation powers (as shown in Figure 3(e) of the main text). In fact, when the EQE is estimated from the higher responsivity value (obtained with 100 nW) it can reach $2{\cdot}10^5$ %.

**Table S3:** EQE for different wavelengths for main text device

| Wavelength [nm] | $I_{ph}$ [nA] | Responsivity [A/W] | EQE [%] | EQE max* [%] |
|---|---|---|---|---|
| 532 | 85 | 58 | 13600 | - |
| 640 | 44 | 30 | 5800 | 200000 |
| 808 | 28 | 19 | 2900 | - |
| 885 | 19 | 13 | 1800 | - |
| 940 | 17 | 12 | 1600 | - |

* This value was obtained using 100 nW laser excitation power

## 6. Responsivity *vs.* gate voltage

In Figure S5 we show the responsivity as a function of backgate voltage. The responsivity here was calculated from the data presented in Figure 3(a) of the main text. The responsivity of the device steadily increases with backgate voltage.

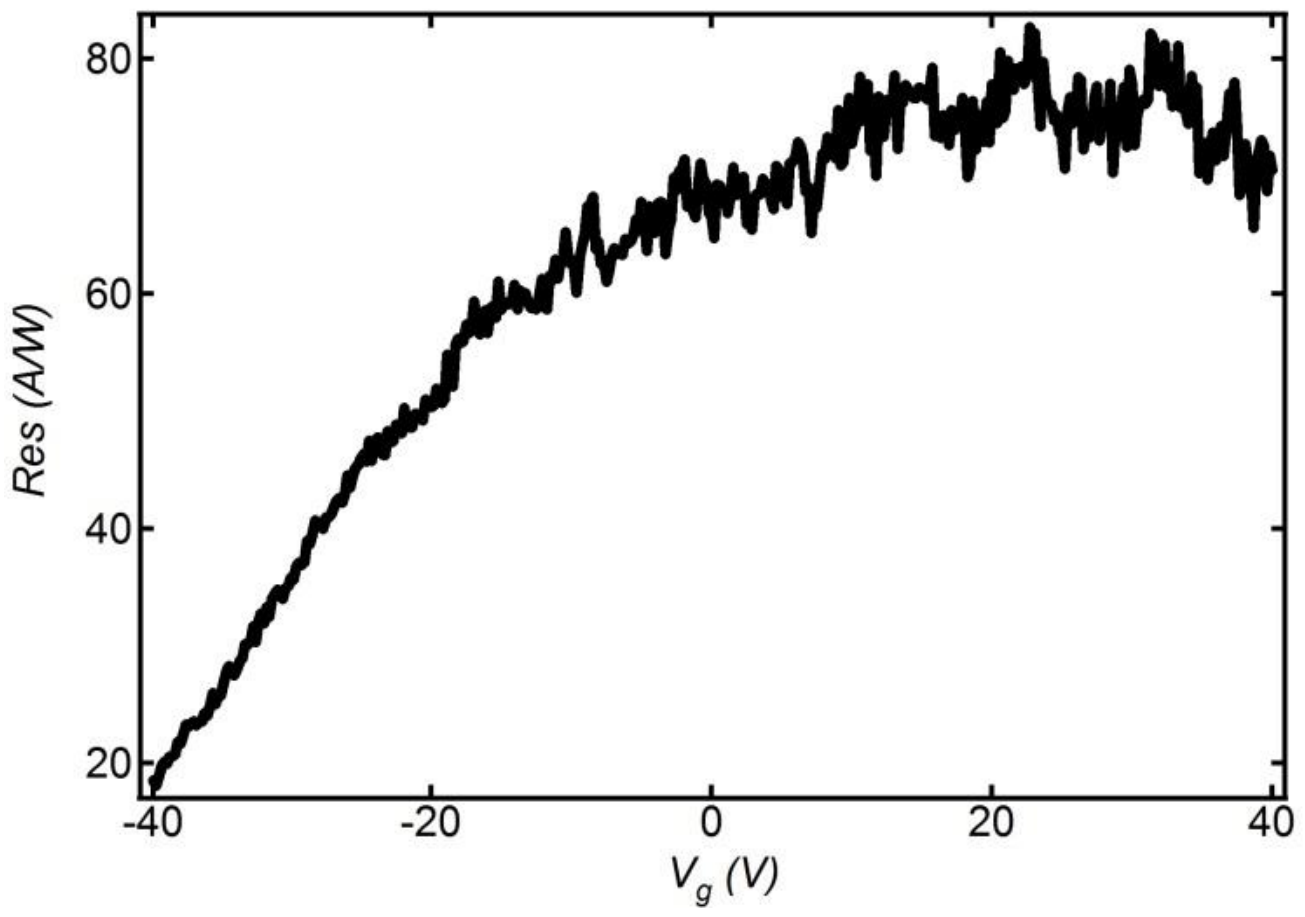

**Figure S5:** Responsivity *vs*. backgate voltage calculated from the main text data in Figure 3(a).